**This preprint has now been published with smaller modifications by Nature Scientific Data.**

It should be cited as:

It is found at

# LUND-PROBE – LUND Prostate Radiotherapy Open Benchmarking and Evaluation dataset

Viktor Rogowski[1,2], Lars E Olsson[1,3], Jonas Scherman[1], Emilia Persson[1,3], Mustafa Kadhim[1,2], Sacha af Wetterstedt[1,3], Adalsteinn Gunnlaugsson[4], Martin P. Nilsson[4], Nandor Vass[4], Mathieu Moreau[4], Maria Gebre Medhin[4,5], Sven Bäck[1,2], Per Munck af Rosenschöld[1,2], Silke Engelholm[4], Christian Jamtheim Gustafsson[1,3]*

1. Radiation Physics, Department of Hematology, Oncology, and Radiation Physics
   Skåne University Hospital, Lund, Sweden
2. Department of Medical Radiation Physics, Clinical Sciences, Lund University, Lund, Sweden
3. Department of Translational Medicine, Medical Radiation Physics
   Lund University, Malmö, Sweden
4. Department of Hematology, Oncology, and Radiation Physics, Skåne University Hospital, Lund, Sweden
5. Department of Clinical Sciences, Lund University, Lund, Sweden

* Corresponding author: Christian.JamtheimGustafsson@skane.se

**Abstract**

Radiotherapy treatment for prostate cancer relies on computed tomography (CT) and/or magnetic resonance imaging (MRI) for segmentation of target volumes and organs at risk (OARs). Manual segmentation of these volumes is regarded as the gold standard for ground truth in machine learning applications but to acquire such data is tedious and time-consuming. A publicly available clinical dataset is presented, comprising MRI- and synthetic CT (sCT) images, target and OARs segmentations, and radiotherapy dose distributions for 432 prostate cancer patients treated with MRI-guided radiotherapy. An extended dataset with 35 patients is also included, with the addition of deep learning (DL)-generated segmentations, DL segmentation uncertainty maps, and DL segmentations manually adjusted by four radiation oncologists. The publication of these resources aims to aid research within the fields of automated radiotherapy treatment planning, segmentation, inter-observer analyses, and DL model uncertainty investigation. The dataset is hosted on the AIDA Data Hub and offers a free-to-use resource for the scientific community, valuable for the advancement of medical imaging and prostate cancer radiotherapy research.

## Background & Summary

Prostate cancer is the second most common diagnosed malignancy among men worldwide and is the fifth leading cause of cancer mortality in men with cancer (1), and requires precise diagnostic and treatment strategies. Prostate cancer can be treated with surgery, hormone therapy, chemotherapy, and/or radiotherapy. The radiotherapy treatment planning process includes several steps, and many steps can be considered labor intensive. Three-dimensional (3D) medical imaging of the patient, with either computed tomography (CT) or magnetic resonance imaging (MRI), or both, is used to define the treatment planning geometry and patient positioning during treatment. These 3D images are annotated through segmentation, where the target of the radiotherapy is referred to as the clinical target volume (CTV), and is manually defined by radiation oncologists. Segmentation of organs-at-risk (OARs) are needed in the treatment planning process to allow for sparing of radiation dose to healthy tissues. A treatment plan is subsequently created and optimized using established clinical dose-volume criteria for the targets and the OARs. This treatment plan will determine how the linear accelerator should deliver the radiotherapy treatment to the specific patient.

Radiotherapy treatment is commonly fractionated, meaning the patient receives multiple doses of radiation over several sessions. A common fractionation for intermediate and high-risk prostate cancer is the ultra-hypo treatment framework. In this regimen, a total radiation dose of 42.7 Gy is delivered in seven fractions to the prostate, over a time span of around three weeks (2).

Prior to the start of radiotherapy, imaging and planning for prostate cancer patients, positioning markers, often referred to as fiducial or implanted markers, are commonly inserted into the prostate. These serve as reference points for image registration, patient positioning before treatment delivery and monitoring during treatment delivery. It is of importance that the prescribed radiation dose is accurately delivered to the target volume as specified in the treatment plan. The fiducial markers are used as a proxy for the location of the prostate and are imaged with the built in X-ray acquisition hardware on the linear accelerator. In this dataset, each patient had three gold fiducial markers inserted into the prostate. The fiducials were cylindrical in shape, measuring 5 mm in length and 1 mm in diameter.

Conventionally, the medical images used for radiotherapy treatment planning originates from a combination of CT and MRI, where CT scans facilitate radiation dose calculation during the optimization of the treatment plan, while MRI provides superior soft tissue contrast and allows for improved segmentation of the target and OARs. CT and MR images can be acquired on the same day or on different days. Either way, anatomical differences such as variation in bladder and rectum filling, may be present and can cause problems when the radiation oncologist defines the ground truth anatomical geometry. Image registration has conventionally been used to enable simultaneous use of CT- and MR-images in the target segmentation process, however this is associated with errors that systematically might propagate through the radiotherapy workflow (3-5). It can therefore be beneficial to use MR images alone for treatment planning. However, MRI cannot be directly used for dose calculation and must therefore be converted to synthetic CT (sCT) images. This is referred to as an MRI-only workflow or MRI-guided treatment planning.

Research on MRI-only workflows has been active for almost two decades and several commercial solutions for creating sCT are available. One of the commercial solutions is MriPlanner from Spectronic Medical (Helsingborg, Sweden), which has been validated and prospectively studied both for prostate (6, 7) and brain (8, 9) radiotherapy. Since 2020, this solution has been used in clinical routine to treat prostate cancer with ultra-hypofractionated radiotherapy at Skåne University Hospital, and in 2023 the routine was clinically implemented for treatment of brain gliomas.

While target and OAR segmentation has traditionally been performed manually, the use of commercial machine learning (ML) models, particularly those based on deep learning (DL), are becoming increasingly integrated into clinical practice for segmenting OARs. Models built on DL have been shown to decrease both the segmentation time and the inter-observer variability (10-12). Despite these advancements, current commercial DL models generate segmentation masks without information on the DL uncertainty. Several methods have been proposed to calculate such uncertainties for radiotherapy DL applications (13), but the clinical implications for end users remain unclear.

In a current study, performed by our research group, four radiation oncologists were asked to rate and edit target and OAR segmentations for 35 prostate cancer patients that were produced by an in-house developed DL model. The rating and editing were performed with and without access to DL segmentation uncertainty information. This approach allowed evaluation of the impact of DL uncertainty on clinical decision-making. It was shown that presenting DL uncertainty information to radiation oncologists could influence their decision-making, perception, and trust in the DL segmentations. This study remains unpublished at the time of writing.

Development and validation of DL models requires access to high quality medical data with consistent segmentations. Several public datasets containing medical images and data for prostate and prostate cancer are available. In a review by Sunoqrot, Saha (14) from 2022, where such datasets were reviewed, it was stated that 3369 multi-vendor prostate MRI cases were available in open datasets, collected 2003-2021. 412 cases were collected for anatomical segmentation tasks, whereas the remainder were collected for prostate cancer detection and/or classification. Since the publication of the work by Sunoqrot, Saha (14) additional public prostate datasets such as PROMISE (15) , Cancer-Net PCa-Data (16), FastMRI Prostate Dataset (17), and ProstateZones (18) have been released.

However, to our knowledge, none of the above public prostate datasets have been acquired in a radiotherapy setting. In 2023 the TROG 15.01 SPARK dataset (19) was released and included radiotherapy planning CT images, treatment plans, dose distribution, fiducial marker annotations, and clinical segmentations for 48 patients. The dataset did not include any MR images and the small size of the dataset might limit the usage for DL model development. In summary, there currently does not exist any publicly available datasets that include medical images, segmentations, and dose distributions alongside multi-observer segmentation. Furthermore, there are no public datasets offering DL segmentation uncertainty information in addition to showcasing the impact on organ segmentation when DL uncertainty information is available to the radiation oncologists. The aim of this work is to present, describe, and provide a publicly available radiotherapy dataset consisting of MR images, synthetic CT images, targets and multiple OAR segmentations, and radiotherapy dose distributions. The base part of the dataset consists of clinical data from 432 MRI-only radiotherapy prostate cancer patients. Moreover, an extended cohort of 35 patients is also included with DL based segmentations of the prostate and rectum together with associated DL uncertainty calculations in addition to the same features as the base cohort. Also, manually edited segmentations of these organs are included where the DL uncertainty both have and have not been presented to the four involved oncologists, to facilitate inter-observer analysis. This data thereby constitutes a complete radiotherapy dataset that can be used by the community for a multitude of different objectives and tasks, including development of novel DL models and advancing prostate cancer radiotherapy research.

## Methods

This dataset comprises clinical data from 432 patients treated with ultra-hypo fractionated prostate radiotherapy using an MRI-only treatment workflow, referred to as the base part of the dataset. The dataset also holds clinical data from 35 patients, for which additional data is included (see below). This is referred to as the extended part of the dataset. Reference to the "dataset" in this paper encompasses both the base and the extended part (432+35 patients). Selection criteria for extracting the data was the following: inclusion into the MRI-only treatment workflow, no hip metal implants, existence of exactly one radiotherapy MRI scan session, exactly one DICOM treatment plan, exactly one DICOM structure set for target and OAR, and exactly one DICOM structure set for fiducial markers. All patients were treated at Skåne University Hospital, Lund, Sweden. A hierarchical overview of the dataset and its content is provided in Figure 1.

All patients in the dataset were prescribed a radiotherapy fractionation scheme of 42.7 Gy in 7 treatment fractions to the prostate gland, excluding seminal vesicles. Radiotherapy planning was performed using the treatment planning system Eclipse (version 15.6, Varian Medical Systems, Palo Alto, USA) using 6 MV or 10 MV flattening filter free (FFF) delivery. Treatment was delivered using single or dual volumetric modulated arc therapy (VMAT) technique using Varian TrueBeam linear accelerators (Varian Medical Systems). Ethical approval for this study was provided by the regional ethics review board in Lund, diary number 2013/742 and by the national ethics review authority, diary number 2024-01720-02. Images, segmentations, and dose data were originally stored in DICOM format, and subsequently anonymized and converted to NIfTI format (20). DICOM structure segmentations were converted to NIfTI using dcmrtstruct2nii (21) and read using Simple ITK (22) in Python.

**MR image acquisition**

The MRI data in this dataset were acquired using a GE Healthcare 3T MRI scanner (GE Healthcare, Chicago, USA). The dataset includes MRI T2-weighted (T2w) images used for prostate target and OAR segmentations utilized during an MRI-only radiotherapy workflow (6). The T2w image volume comprised a large field-of-view (FOV) of 44.8-48 cm and was acquired in the transversal plane to encompass the whole pelvic contour. MRI T2w data was acquired with a 2-dimensional (2D) based fast spin echo acquisition with an in-plane resolution between 0.6x0.7 to 0.8x0.9 mm and a slice thickness of 2.5 mm. The output reconstructed in-plane voxel size in the dataset is slightly smaller than the MRI acquisition voxel size due to up sampling of the image matrix in the MRI reconstruction software, see section Data Records for further details.

Due to continuous upgrades of the MRI scanner hardware and image reconstruction techniques over the years, enabling improved and faster image acquisition, two versions of the MR image acquisition protocol were defined, and the dataset therefore contains data from two different versions of the MRI acquisition protocol. 325 patients out of the 432 patients in the base part were acquired with the MRI acquisition protocol described by Persson, Jamtheim Gustafsson (6) and is here referred to as the old acquisition protocol (oldAcq). A total of 109 patients out of the 432 patients in the base part were acquired with the MRI acquisition protocol defined in the supplement of Olsson, Af Wetterstedt (23), which is referred to as the new acquisition protocol (newAcq). For the patients in the extended part, all MRI data were acquired with the new MRI acquisition protocol. The largest and most substantial differences between the oldAcq and newAcq protocols are that the newAcq is less sensitive to patient motion in the anterior-posterior direction and includes GE Healthcare AirRecon (GE Healthcare, Chicago, USA) in the image reconstruction to reduce Gibbs artefacts and image noise (24).

### Synthetic CT generation

The sCTs were generated from the T2w MRI volume using the MriPlanner (Spectronic Medical AB, Helsingborg, Sweden) software versions 2.3 and 2.4 (25). Some of the sCTs in the dataset contain a warning text message that was inserted onto the image stating "Synthetic CT. Not for diagnostics". This was initially included to avoid misuse of sCT as an ordinary CT for radiology diagnostic tasks. The sCT voxel size was thereby not the same as the MRI data voxel size. A resampled version, using bspline interpolation, of the sCT has therefore been provided to match the MRI voxel size and geometry, see Data Records for details. In the resampled version of the sCT, the warning text message has been removed.

The center of mass positions for each of the three inserted prostate fiducial markers were manually annotated in Eclipse by the MRI technicians, based on clinical routines utilizing Multi Echo Gradient Echo imaging (26, 27). A high-density sphere with 2.5 mm radius was thereafter created at each center of mass point in the sCT image by the MriPlanner software. Spatial locations for these fiducials are also available in this dataset as DICOM coordinate points in the MRI geometry space. Additionally, the fiducial marker location is provided in the dataset as a binary segmentation mask in the MRI geometry, where each fiducial center of mass defines the center of a spherical object with 3 mm radius. This segmentation mask thereby defines three spherical objects. Please observe that these spherical objects are not the same size as the high-density spherical objects created in the sCT.

### Clinical structure segmentations

The entire prostate gland was defined as the clinical target volume (CTV), from here on referred to as the prostate CTV, was manually segmented by radiation oncologists according to clinical routine. A 7 mm isotropic margin accounting for setup uncertainties in the radiotherapy treatment processes was added to generate the planning target volume (PTV). Segmentation of the prostate CTV and rectum followed ESTRO guidelines (28). OARs such as the rectum, genitalia and penile bulb were also manually segmented by radiation oncologists. The bladder and femoral heads were initially segmented using a DL module from the MriPlanner software but were subsequently verified and edited as necessary by the dosimetrists during the treatment planning process. The body contour was automatically created by Eclipse treatment planning system, reviewed, and manually edited if needed. The name standard provided in (29, 30) was used for labeling the targets and OAR and this was propagated to the respective file names in this dataset. Provided segmentations for the radiotherapy treatment plan were approved in the clinical review process and the plan was thereafter used for the delivery of the treatment. All target and OAR segmentations in the dataset were also resampled to the MRI volume geometry, as the sCT and the MR image volume did not have the same matrix or voxel size. This ensured accurate alignment and consistency for further use of the dataset image and segmentation data.

**Dose distribution data**

The dose distribution from the optimized treatment plan was calculated with the Analytical Anisotropic Algorithm (AAA, algorithm version 15.6) using the treatment planning system Eclipse. For each patient in the dataset, the dose distribution is included with the original dose voxel size of 2.5x2.5x2.5 mm or 1x1x0.83 mm in a separate file. Dose distributions were also interpolated and resampled to the same voxel size as the sCT, using linear interpolation. This was performed to provide a voxel-to-voxel oriented pair of sCT and dose data in the dataset. For the same reason, the dose distribution was also interpolated and resampled using the same settings as above to the MRI geometry. All dose distributions in the dataset were rescaled to the unit Gray (Gy) using the Dose Grid Scaling DICOM attribute.

**Description of data in extended part**

The extended part of the dataset contains patient data with the same type of data as the base part but extended to include additional features: DL generated segmentations, DL uncertainty maps, and DL segmentations reviewed and adjusted by four radiation oncologists. This data has been generated and utilized in another study investigating DL segmentation uncertainty, referred to as the uncertainty study. The following section will describe the methodology used to generate these extended features.

**Deep learning segmentation generation**

Separate DL models were developed for segmentation of the prostate CTV and the rectum as an OAR on the large FOV T2w MR images using the nnUNet framework (two models in total, version 2) (31). Model training data utilized the 432 patients in the base dataset, employing a 10-fold cross validation (CV) approach instead of the default nnUNet 5-fold CV. By default, nnUNet relies on model ensemble and calculates the average predictions from all trained folds to produce the final segmentation. This approach leverages the variability between models trained on different data splits, improving robustness and accuracy by combining their outputs. The average validation Dice score of the prostate CTV and the rectum models, per fold, was 0.89-0.90 (n=10 CV-folds) and 0.86-0.89 (n=10 CV-folds), respectively.

**Generation of DL uncertainty map**

The extended part of the dataset was used for DL model inference and served as an independent test dataset. To estimate the DL uncertainty for the prostate CTV and rectum model, the SoftMax values (ranging from 0 to 1) by each of the ten CV-models after inference were stored. Next, the SoftMax standard deviation was calculated voxel-wise for each patient within the prostate CTV and rectum regions. The extended dataset includes the final DL segmentation generated by the model ensemble for each patient as well as the individual segmentation produced by each of the ten CV model folds and the calculated uncertainty map. This comprehensive inclusion allows for detailed analyses of both the segmentation performance and the associated uncertainty.

**Impact of uncertainty and generation of edited DL segmentations**

The uncertainty study was designed to evaluate the impact of uncertainty in DL generated segmentations of the prostate CTV and rectum on oncologists. The study was conducted in two steps, Step 1: Four oncologists were tasked with editing the DL generated segmentations without access to segmentation uncertainty information. Step 2: Four weeks later, the same oncologists repeated the same task as in step 1, for the same set of patients, this time with the uncertainty information presented as an image color overlay (heatmap). Edited DL segmentations for prostate CTV and rectum, for each patient, from both steps are available for all four oncologists in the extended part of the dataset.

## Data Records

The dataset is hosted on the AIDA Data Hub (https://datahub.aida.scilifelab.se/10.23698/aida/lund-probe) where a digital request can be made to download the dataset. All data is provided in compressed NIfTI format (.nii.gz) as it allows for geometry metadata to be included, while removing all other sensitive patient and imaging metadata. The full contents of the dataset are described for every file type in https://github.com/jamtheim/LUND-PROBE/blob/main/Data_table.pdf where the relative location of the data within the dataset is specified.

Not all patients in the dataset had a complete set of radiotherapy segmentations, as defined in https://github.com/jamtheim/LUND-PROBE/blob/main/Data_table.pdf. In the base part of the dataset with 432 patients, 29 patients did not have the penile bulb segmented, and 24 patients did not have the genitalia segmented. A file called missingStructures_basePart.json in the dataset defines this missing data. In the extended part with 35 patients, one patient had a missing PenileBulb segmentation, defined in the dataset by the file missingStructures_extendedPart.json.

From a clinical radiotherapy planning perspective, it is desirable that each segmentation is defined in every image slice of the encompassing organ volume. When compiling this dataset, a data quality assurance step for this purpose was implemented by analyzing whether any segmentations contained empty slices, utilizing a 26-connected component analysis. Segmentations that deviated from this standard, such as those with slice gaps in the segmentation, were identified and added to the files StructuresMoreThan1ConComp_basePart.txt and StructuresMoreThan1ConComp_extendedPart.txt for the base and extended part of the dataset, respectively. Such information can be of importance e.g., when using segmentation data for ML training and evaluation. In total, 21 segmentations in the dataset were identified to have one or several missing slices in the encompassing organ volume. The most affected OAR was the bladder.

The MR images from the oldAcq and newAct acquisition protocols had slightly different reconstructed voxel resolution where the oldAcq had a voxel resolution of 0.4375x0.4375x2.5 mm and the newAct acquisition protocol had 0.4688x0.4688x2.5 mm. Deviations to the above were observed and we therefore recorded the matrix and voxel size for sCT and MRI for each patient in the files patGeometryInformation_basePart.csv and patGeometryInformation_extendedPart.csv, included in the dataset. These files also contain the matrix size and voxel size for the original dose distribution file. To facilitate a voxel-to-voxel correspondence between the sCT and MR images, a registration and resampling of the sCT to the MRI geometry was performed for each patient and saved with file name image_reg2MRI.nii.gz.

For the 35 patients in the extended part of the dataset, which were also used in the mentioned uncertainty study the nnUNet segmentations and uncertainty maps are provided in the folder “nnUNet” accompanied by the individual segmentations for each CV-fold in the subfolder “folds”. The edited DL segmentation in Step 1 and Step 2 is saved in the folder “observerData” where the file prefix also defines the oncologist who performed the adjustment of the DL segmentation. The four oncologists are referred to as obsB, obsC, obsD and obsE.

## Technical Validation

To ensure accurate alignment between the segmentations and the corresponding patient image volume and image geometries, key metadata in the NIfTI headers were verified for consistency between the segmentation files and the sCT and MR image volumes. Similar verification was performed for the dose distribution and DL uncertainty files. For the dose file, the unit used is Gy and its maximum value was verified to be in line with the dose prescription. Organ segmentations together with a fiducial marker are visualized and overlayed on MRI and sCT image volume in Figure 2 (a-b). An sCT image with overlayed dose distribution is also presented in Figure 2c. MRI volume together with multiple oncologists' segmentation for prostate CTV and an example of the DL uncertainty map is seen in Figure 3. All targets and OAR available in the dataset are visualized in 3D with respect to patient orientation in Figure 4.

## Usage Notes

The provided dataset comprises a unique collection of radiotherapy data for prostate cancer treatment which includes images, segmentations, dose distributions, original DL segmentations, oncologists edited DL segmentations, DL-uncertainty, and inter-observer data. This dataset may facilitate the development of several DL applications, such as DL segmentation, DL synthetic image generation, DL dose prediction, impact analysis of DL uncertainty and inter-observer differences. The completeness of the data from a clinical radiotherapy perspective also allows it to be used in external benchmarks for ML models developed on other data. The high-quality MRI volumes of the pelvis allow the data to be utilized in additional segmentation studies, similar to the approach of Holmlund, Simko (18) who used the available PROSTATEx dataset (32). The publicly available dataset can be used for commercial applications.

## Code Availability

Python code for loading, visualizing, and handling the data is provided in a Jupyter notebook, located at the GitHub repository https://github.com/jamtheim/LUND-PROBE.

## Acknowledgements

This work was supported by Skåne University Hospital and governmental funding of clinical research within the National Health Service.

## Competing Interests

The authors declare no competing interests.

## Figures

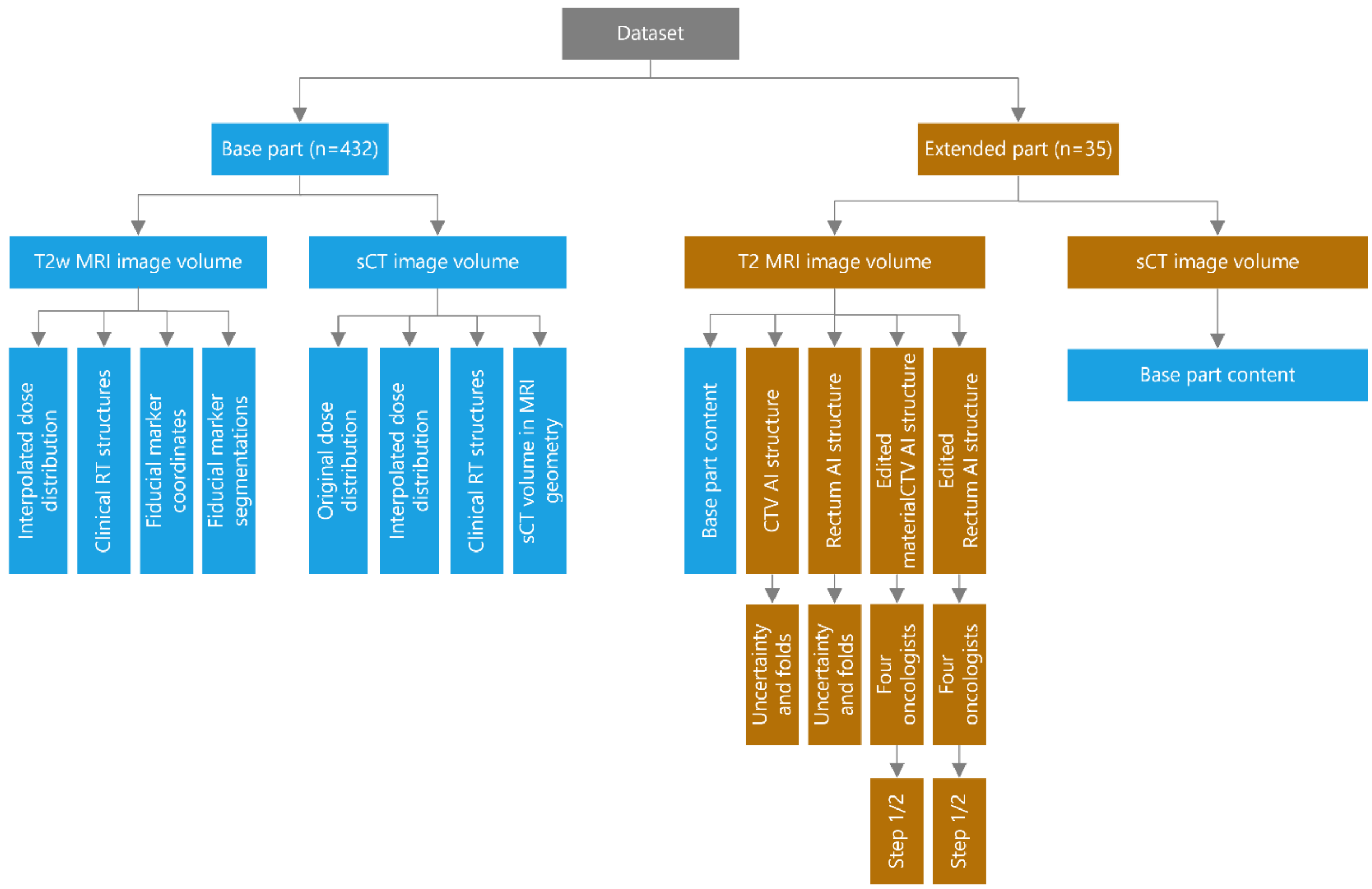

Figure 1. Data overview. A hierarchical overview of the dataset and its content for base part (n=432, blue color, left) and extended part (n=35, gold color, right). The content in the base part was included in the extended part for each patient.

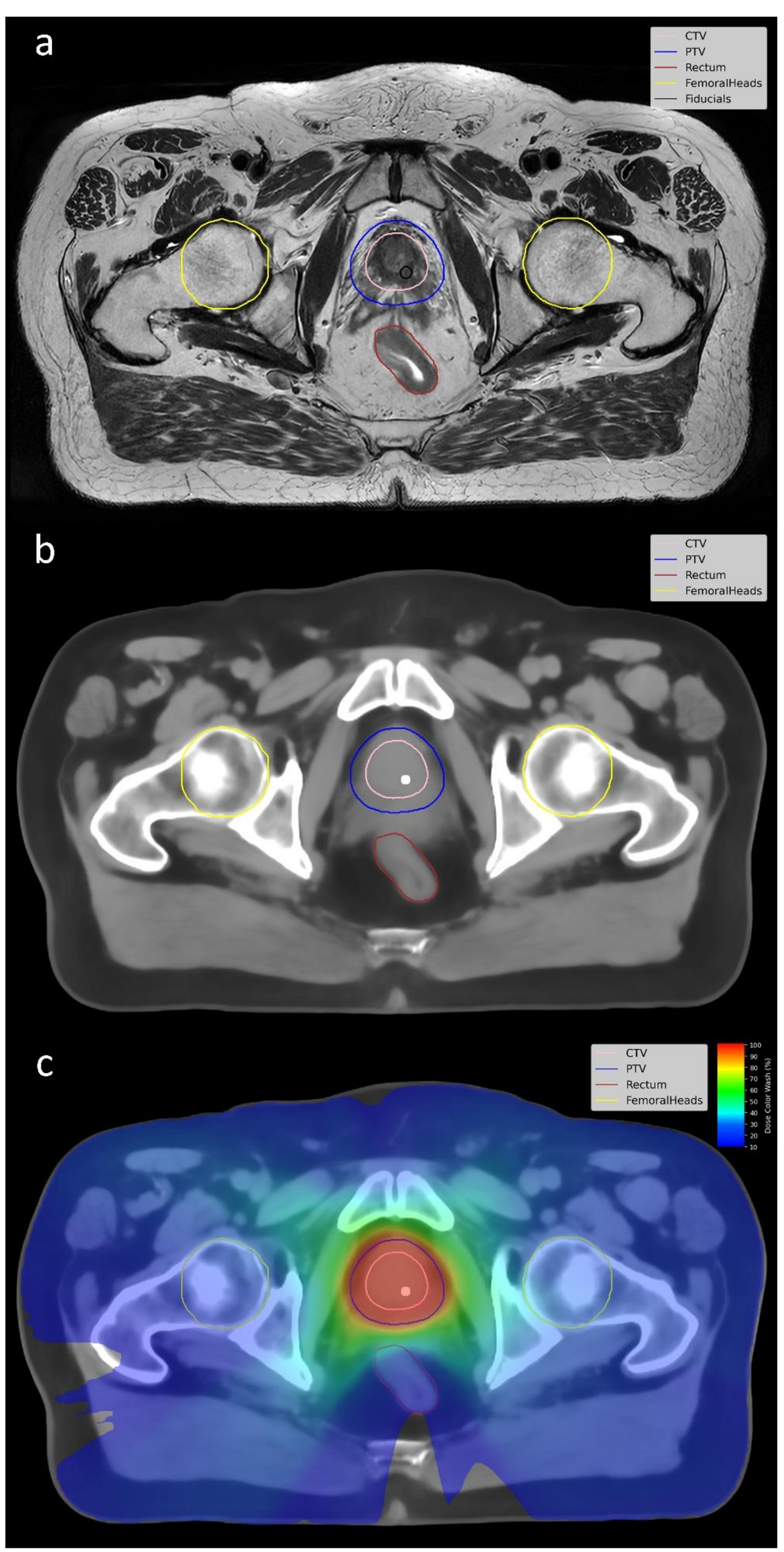

Figure 2. Image data types in the dataset. a) MRI T2w volume with clinical radiotherapy segmentations overlayed. The center of the fiducial marker delineation in black shows the defined center of mass for one of the fiducial markers inserted into the prostate. b) Synthetic CT (sCT) volume with radiotherapy segmentations overlayed. The sCT was created from the MRI T2w images in a) using the Spectronic MriPlanner software. The high-density sphere in the prostate, seen as a high intensity object, is placed at the fiducial marker center of mass point defined in the MRI T2w volume. c) Synthetic CT (sCT) with radiotherapy segmentations and overlayed dose distribution, see dose color wash legend to the right where 100% equals 42.7 Gy. It is the same patient and corresponding image slice in a), b), and c).

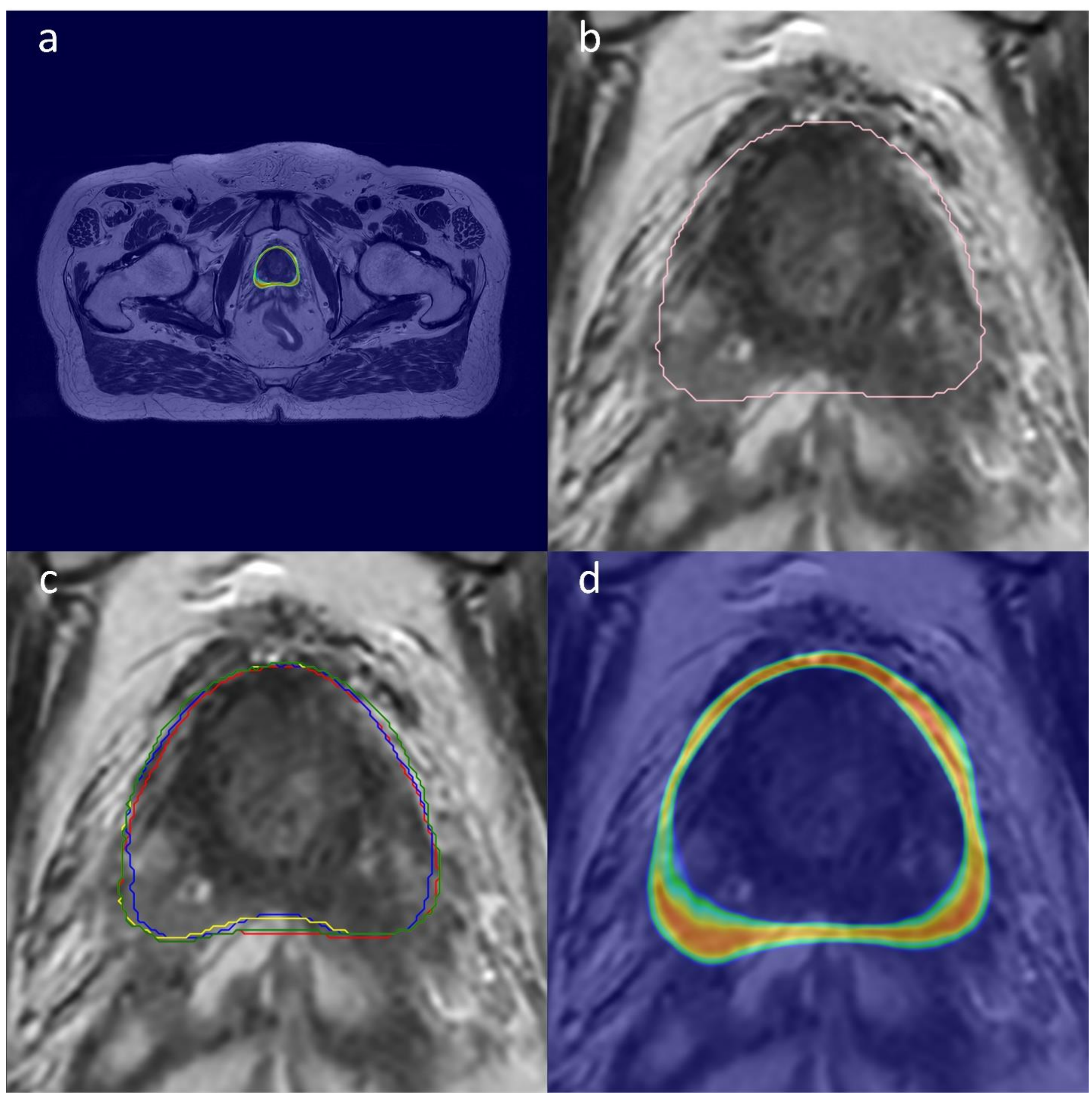

Figure 3. Prostate segmentations. a) MRI T2w image slice with prostate CTV deep learning segmentation uncertainty map overlaid. b) T2w MRI zoomed in with prostate deep learning CTV segmentation. c) Four different oncologist's individual prostate CTV segmentations on zoomed in T2w MRI, visualized in separate colors. d) deep learning prostate CTV segmentation uncertainty map visualized in color, zoomed in from a). Corresponding data for rectum is available in the dataset.

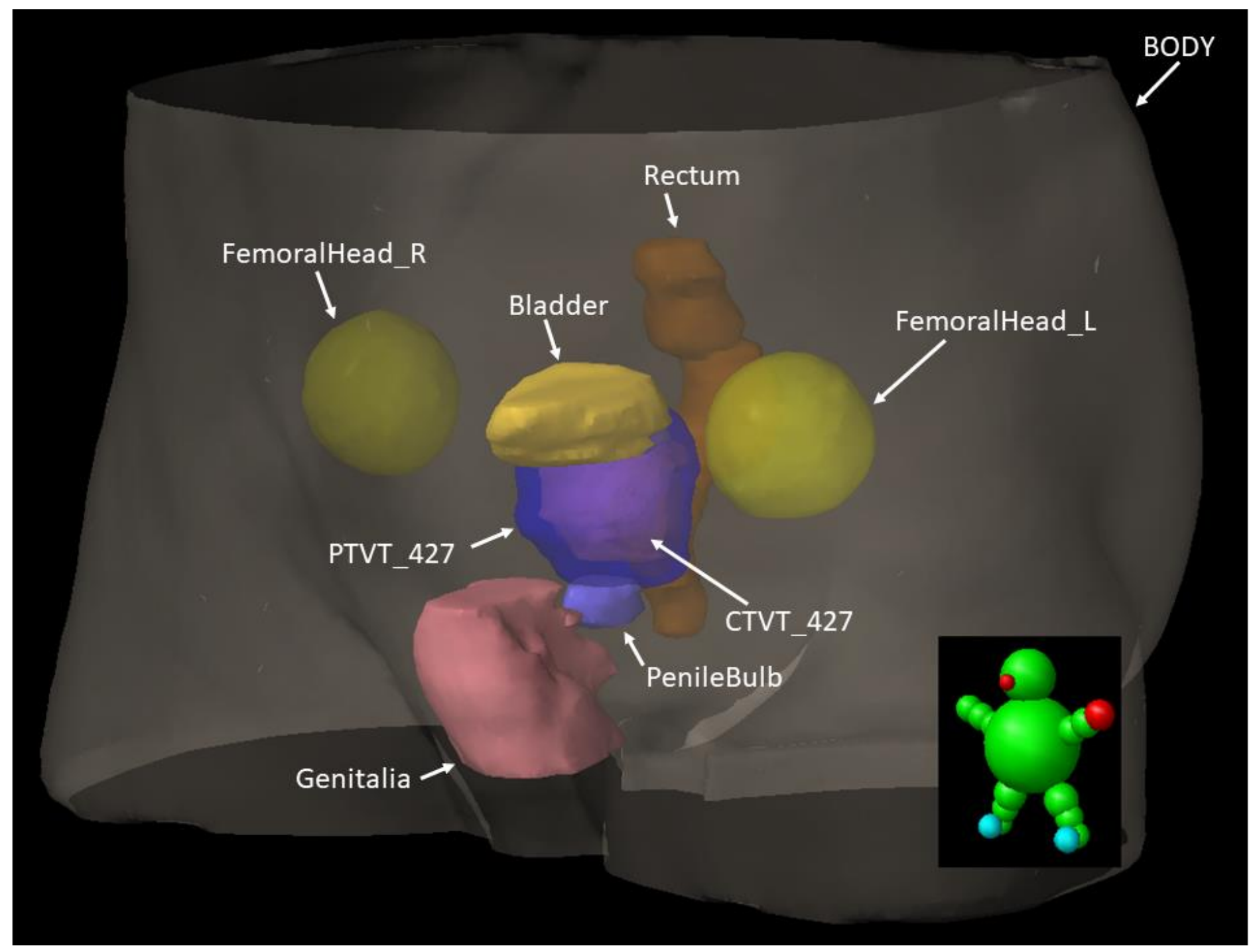

Figure 4. Pelvis segmentation overview. One patient volume oriented as demonstrated by the green model. Overlayed are the available prostate targets where PTVT_427 (dark blue) encompasses the prostate CTVT_427 (purple) together with the Bladder (yellow), left (FemoralHead_L) and right (FemoralHead_R) femoral heads (green-yellow), PenileBulb (light blue), Rectum (brown) and Genitalia (red), all available with respective name in the dataset. The BODY segmentation, encompassing the whole scanned patient volume, is the largest volume. Fiducial marker delineations are not shown.